\documentstyle[pre,aps,floats]{revtex}
\input epsf
\begin{document}

\twocolumn[
\title{\bf Scaling of Reaction Zones in the $\protect\bbox{A+B\rightarrow
\emptyset}$\/ Diffusion-Limited Reaction}
\author{Benjamin P. Lee\cite{IPST} and John Cardy}
\address{Theoretical Physics, University of Oxford, 1 Keble Road, Oxford
OX1 3NP, United Kingdom.}
\date{Preprint: \today}
\widetext\leftskip=0.10753\textwidth \rightskip\leftskip
\begin{abstract}
We study reaction zones in three different versions of the
$A+B\rightarrow\emptyset$ system. For a steady state
formed by opposing currents of $A$ and $B$ particles we derive scaling
behavior via renormalization group analysis.  By use of a previously
developed analogy, these results are extended to  the
time-dependent case of an initially segregated system.  We also
consider an initially mixed system, which forms reaction zones
for dimension $d<4$.  In this case an extension of the steady-state
analogy gives scaling results characterized by new exponents.
\end{abstract}

\bgroup\draft
\egroup
\maketitle]

The reaction-diffusion process $A+B\rightarrow\emptyset$ has received
much attention since the work of Toussaint and Wilczek \cite{TW}, in
which it was demonstrated that density fluctuations are important
when the dimensionality $d<4$.  Attempts to incorporate the
fluctuation effects, including the application of renormalization
group (RG) methods, have proven successful for one-species reactions
\cite{Peliti,DS,Lee}.  However, the physics of the two-species reaction
is somewhat different, as the number difference of $A$ and $B$
particles is locally conserved.  Nonetheless, the main results of the
RG analysis may be extended to this case \cite{Leeii,LC}.

In the present work we are concerned with situations in which the $A$
and $B$ particles are well segregated, so that reactions are then
confined to reaction zones, on the boundaries between the $A$- and
$B$-dominated regions.  The segregation may be a consequence of the
initial conditions, or arise asymptotically from a mixed initial
state when $d<4$.  Both of these cases may be compared to that of
a steady state formed by directing steady, uniform currents of $A$
and $B$ particles towards each other.  In all three cases the
structures of the reaction zones are very similar.

Reaction zones were first studied by G\'alfi and R\'acz \cite{GR}
in the context of a system with segregated initial conditions.
Their analysis is essentially a mean-field result, as we will show,
which holds asymptotically for $d>2$.  Later, Cornell and Droz
treated the case of $d<2$ by using scaling arguments motivated by a
renormalization group picture \cite{CD}.  Also, reaction zones in
initially homogeneous systems were studied by Leyvraz and Redner
\cite{LR,Leyvraz}.  In this paper we use exact RG methods both to
confirm previous results and to derive new ones for the exponents
which characterize the reaction zones.

The model we shall use for the $A+B\rightarrow\emptyset$ reaction
is one of particles undergoing continuous-time random walks on a
hypercubic lattice. We consider general, non-zero values of the
diffusion constants $D_A$ and $D_B$.  If an $A$ and a $B$ particle
are together on a lattice site, then they annihilate with some
characteristic reaction rate $\lambda$.  In the field-theoretic
approach it is convenient to allow multiple occupancy of lattice
sites, regardless of particle type, but this is not expected to alter
the universal properties.

First we consider the steady state reached when equal currents $J$
of $A$ and $B$ particles are directed towards each other.  The
average densities then vary only in the direction of the currents.
In this case the width of the reaction zone goes as a power of $J$,
as $J\rightarrow 0$,
\begin{equation}\label{wJ}
w\sim\cases{J^{-1/(d+1)}&$d\le 2$\cr J^{-1/3}& $d> 2$.\cr}
\end{equation}
The typical nearest-neighbor distance in the reaction zone
$\ell_{\rm rz}$ scales as
\begin{equation}\label{lJ}
\ell_{\rm rz}\sim\cases{J^{-1/(d+1)}&$d\le 2$\cr J^{-2/3d}&
$d>2$.\cr}
\end{equation}

The second example we consider is that of initially segregated
particles.  Consider a system prepared with only $A$ particles in
the region $x<0$ and only $B$ particles for $x>0$.  The behavior in
the reaction zone again scales, now with respect to time, with
$w\sim t^\alpha$ and $\ell_{\rm rz}\sim t^\gamma$.  The densities
in the reaction zone, where $|x-x_c|\lesssim w$, have the scaling
form
\begin{equation}\label{scale:a}
\langle a\rangle,\langle b\rangle\sim t^{-\gamma d}F_{a,b}\left(
x-x_c\over t^\alpha\right)
\end{equation}
where $x_c\propto t^{1/2}$ is the reaction zone center.  The angular
brackets will be defined precisely below.  The reaction rate $R$
also has the scaling form
\begin{equation}\label{scale:R}
R(x,t)\sim t^{-\beta}G\left(x-x_c\over t^\alpha\right).
\end{equation}
The values of the exponents are given in table \ref{tb}.

\begin{table}
\caption{The exponents, as defined in the text, for the scaling
behavior in the cases of segregated and homogeneous initial
conditions.} \label{tb}
\begin{tabular}{lcccc}
 & & $\alpha$ &$\beta$&$\gamma$ \\ [.05cm]\hline
\rule[-0.3cm]{0cm}{0.8cm}
Segregated  &$d\le 2$ & ${\displaystyle 1\over\displaystyle 2(d+1)}$
& ${\displaystyle d+2\over\displaystyle 2(d+1)}$
&${\displaystyle 1\over\displaystyle 2(d+1)}$ \\[.3cm]
 &$d>2$  & ${\displaystyle 1\over\displaystyle 6}$ & ${\displaystyle
2\over\displaystyle 3}$ & ${\displaystyle 1\over\displaystyle 3d}$
\\[.2cm] Homogeneous &$d\le 2$ & ${\displaystyle d+2\over\displaystyle
4(d+1)}$ & ${\displaystyle (d+2)^2\over\displaystyle 4(d+1)}$
& ${\displaystyle d+2\over\displaystyle 4(d+1)}$ \\[.4cm]
 &$2<d<4$  & ${\displaystyle d+2\over\displaystyle 12}$& ${\displaystyle
d+2\over\displaystyle 3}$ & ${\displaystyle d+2\over\displaystyle 6d}$
\\[.2cm]
\end{tabular}
\end{table}

The final example is that of random, homogeneous initial conditions,
with equal densities $n_0$ of $A$ and $B$ particles.  It has been
shown in the mathematical treatment of Bramson and Lebowitz \cite{BL}
that this system exhibits asymptotic segregation for $d<4$.  We
present a simple derivation of this result from an effective field
theory, which is valid for $2<d<4$.  We argue that reaction zones
which result from the segregation can be studied in the same fashion
as in the previous problems.  As a result, we predict the same scaling
forms as given by Eqs.\ (\ref{scale:a}) and (\ref{scale:R}), with new
values for the exponents, which are listed in rows $3$ and $4$ of
table \ref{tb}.

The characteristic nearest-neighbor distance in the reaction zone
$\ell_{\rm rz}$ is found from the scaling form of the densities in
the reaction zone (\ref{scale:a}).  Consider the total number of
particles $N$ in the volume of the reaction zone for a given
interfacial surface area $S_{d-1}$,
\begin{equation}
N\sim S_{d-1}\int_{x_c-w}^{x_c+w}dx\>\langle a(x)+b(x)\rangle
\sim S_{d-1}t^{\alpha-\gamma d}.
\end{equation}
Since the volume of this region is $S_{d-1}t^\alpha$, then the
average volume per particle goes as $t^{\gamma d}$.  If one assumes
that there is just one length scale describing typical
nearest-neighbor distance, both in the direction of the interface
and perpendicular to it, and also for both like and unlike particles,
then this length is given by $\ell_{\rm rz}\sim t^{\gamma}$.
A convenient definition of $\ell_{\rm rz}$, which follows from the
derivation above, is $\ell_{\rm rz}\equiv\langle a(x_c)
\rangle^{-1/d}$.

The model for $A+B\rightarrow\emptyset$ can be mapped to a field
theory, which is useful for the application of RG methods
\cite{Leeii,PelitiRev}.  We summarize here the results of a study
of this field theory which pertain to the problem at hand.  A more
detailed account will appear in \cite{LC}.  One feature, which is
general to all irreversible reaction-diffusion systems, is that there
are no diagrams which dress the propagator.  This means that there
is no wavefunction renormalization, and consequently no anomalous
dimension for the fields. The significance of this will be noted below.

There is only one coupling constant $\lambda_0$ in the field theory
(in the notation of \cite{Leeii}), which is given in terms of the
original microscopic parameters as $\lambda_0=\lambda h^d/\bar D$,
where $\bar D=(D_A+D_B)/2$ and $h$ is the lattice size.  The coupling
$\lambda_0$ is found to be irrelevant, in the RG sense, for $d>2$,
marginal for $d=2$, and relevant for $d<2$.

In general one can derive from the field theory the equations of motion
\begin{equation}\label{eom}
\partial_t\langle a\rangle=D_A\nabla^2\langle a\rangle-R\qquad
\partial_t\langle b\rangle=D_B\nabla^2\langle b\rangle-R
\end{equation}
where $R=\lambda_0\bar D\langle ab\rangle$ is the reaction rate.  The
angular brackets refer to averages over the stochastic processes of
diffusion and reaction, but not over the initial conditions.  Averages
over the initial conditions will be denoted with a bar:
$\overline{\langle a\rangle}$.  Note that the quantity $\langle a-b
\rangle$ obeys a simple diffusion equation when $D_A=D_B$.

When the $d>2$ the irrelevance of the coupling leads to the asymptotic
result that $R\sim\Gamma\langle a\rangle\langle b\rangle$.  Then the
equations of motion (\ref{eom}) are simply differential equations with
rate constant $\Gamma$, which is the starting point in the
analysis of G\'alfi and R\'acz \cite{GR}. Therefore their results
are applicable for $d>2$.
When $d\le 2$ one finds that the coupling $\lambda_0$ can be renormalized
exactly, allowing us to derive the scaling behavior of the reaction zones.

Consider the steady state reached by imposing currents ${\bf J}_a
=J{\bf\hat x}$ at $x=-L$ and ${\bf J}_b=-J{\bf\hat x}$ at $x=L$, where
$L\gg w$.  From the analysis in \cite{GR} one finds for $d>2$ that
\begin{equation} \label{Rd>2}
R\sim\langle a\rangle\langle b\rangle\sim J^{4/3}f(xJ^{1/3}),
\end{equation}
from which the exponents for $d>2$ given in Eqs.\ (\ref{wJ}) and (\ref{lJ})
follow directly.

For $d\le 2$ one must consider the effects of renormalization.  We
work in units of time where the average diffusion constant $\bar D=1$,
and define the parameter $\delta=(D_A-D_B)/(D_A+D_B)$.  One can show
that $\delta$ is not renormalized, for the same reason that there is
no wave-function renormalization.  In these units
the dimension of the current is $[J]=(\mbox{length})^{-1-d}$.
The width $w$ can depend in the steady state on $\lambda_0$, $J$,
and $\delta$ only.

In the field-theoretic RG, the $\lambda_0$-dependence is traded for
a dependence on a renormalized coupling $\lambda_R$, defined at an
arbitrary length scale $J_0^{-1/(d+1)}$.  We then consider $w$ as
depending on $J$, $\delta$, $J_0$, and the the dimensionless renormalized
coupling $g_R\equiv\lambda_RJ_0^{(d-2)/(d+1)}$.  This then satisfies an
RG equation whose solution is
\begin{equation}\label{CSsol}
w(J,g_r,J_0,\delta)=\biggl({J\over J_0}\biggr)^{-1/(d+1)}w\Bigl(J_0,
\tilde g(J/J_0,g_R),\delta\Bigr)
\end{equation}
where, as $J\rightarrow 0$, the running coupling $\tilde g\rightarrow
g^*=O(2-d)$, {\it independently} of the initial value of $\lambda_0$.
When $d=2$ the running coupling goes as $\tilde g\sim B/\ln|J|$
for small $J$.
The simple form of the overall scale factor is a result of the absence
of wave-function renormalization.

The significance of Eq.\ (\ref{CSsol}) is that, in the limit of small
$J$, the only $J$ dependence on the right-hand side comes from the
overall scale factor.  Therefore any length
will scale asymptotically as $w\sim\ell_{\rm rz}\sim J^{-1/(d+1)}$,
giving the results stated Eqs. (\ref{wJ}) and (\ref{lJ}) for $d\le 2$.
When $d=2$ the running coupling $\tilde g$ is $J$ dependent.  However,
it can be shown that the right-hand side is finite in the limit
$\tilde g\rightarrow 0$ \cite{LC}.  Therefore the leading term for
small $J$ is unchanged, and there are no
logarithmic corrections.  It should be noted that while this result
holds for general $|\delta|<1$,
the limit of $|\delta|\rightarrow 1$ does not necessarily commute with
the asymptotic limit in our analysis, so our results are only for the
case when both species are mobile.

The asymptotic limit of any dimensionful quantity can be found by
the same technique, using only dimensional analysis.  Therefore the
densities in the reaction zone have the scaling form
\begin{equation}
\langle a(x)\rangle,\langle b(x)\rangle\sim J^{d/(d+1)}
F_{a,b}(xJ^{1/(d+1)})
\end{equation}
and the reaction rate
\begin{equation} \label{Rd<2}
R(x)\sim J^{(d+2)/(d+1)}G(xJ^{1/(d+1)}).
\end{equation}
The scaling functions $F_{a,b}$ and $G$ are expected to be universal, up
to metric factors, and should be calculable within a $(2-d)$-expansion.
Note that $R\propto\langle ab\rangle\sim J^{(2-d)/(d+1)}\langle a\rangle
\langle b\rangle$ for small $J$.  This result can be demonstrated
explicitly within the original field theory \cite{LC}.

An equivalent and less formal way to state these results is that there
are only two input parameters with dimension: $J$, and $\lambda_0$.
However, for $d\le 2$, in the asymptotic limit the theory is independent
of the initial value of $\lambda_0$, so the remaining $J$ dependence can be
determined on dimensional grounds.

These results were found by Cornell and Droz \cite{CD}, by use of an
RG-motivated scaling analysis.  Our work provides a quantitative
confirmation of the lack of anomalous dimension and asymptotic lack
of $\lambda_0$ dependence.  They performed simulations for $d=1,2,3$
which are in good agreement with the predictions.

\begin{figure}
\epsfxsize=3in
\centerline{\epsfbox{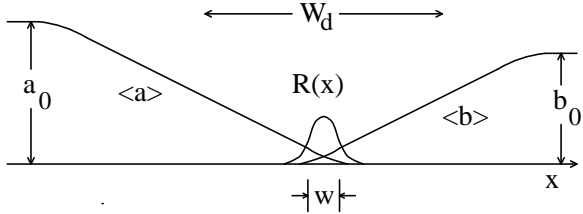}}
\caption{The profile of a reaction zone.}\label{f:rz}
\end{figure}

We now turn to the problem of segregated initial conditions.  At $t=0$
the boundary is at $x=0$, and the $A$ and $B$ particles are randomly
distributed within their regions, with initial densities $a_0$ and $b_0$
respectively.  A profile of the densities at a later time $t$ is
sketched in Fig.\ \ref{f:rz}.  The densities are depleted out to a
range $W_d\sim t^{1/2}$, which is the length over which particles
will have had a chance to diffuse into reaction range.  Provided that
$\alpha<1/2$, which will be verified self-consistently, then
asymptotically $w\ll W_d$.  Consequently one finds in the depletion
regions  $w\ll|x-x_c|\ll W_d$ that $R\approx 0$
and Eq.\ (\ref{eom}) reduces to the diffusion equation.  As a result,
it follows that the density profiles in this depletion region are
linear in $x-x_c$.  The slope can be determined by observing that
the $A$ particle density goes from $a_0$ to zero in a range $W_d$, so
the slope  $-a_0/W_d\sim -t^{-1/2}$.

Implicit in the analysis of G\'alfi and R\'acz \cite{GR} is the
observation that this linear depletion regime
provides the same boundary conditions for the reaction zone as was found
in the steady-state system.  In that case the slope in the linear region
($w\ll |x|<L$) is given by $\partial_x\langle a\rangle=-J$.  Therefore
we can apply
the results obtained from the steady state to the time-dependent reaction
zone by making the scaling substitution $J\sim t^{-1/2}$.  The result
is the scaling forms given by Eqs.\ (\ref{scale:a}) and Eq.\
(\ref{scale:R}) with the exponents shown in table \ref{tb}.
The exponent $\gamma$ for the nearest-neighbor distance $\ell_{\rm rz}$ is,
to our knowledge, a new result.

The exponents $\alpha$ and $\beta$, which describe $w$ and $R$, were
derived in the analysis of Cornell and Droz \cite{CD}.  They find
reasonable agreement with these predictions in $d=1$
simulations, although the asymptotic region is difficult to obtain.
Other simulations in $d=1$ find evidence for multi-scaling \cite{Stanley}.
Since the parameters in these simulations are at the opposite extreme
from the weak coupling expansion implicit in the field-theoretic approach,
it is conceivable that they might fall into a separate universality
class.  However,  more extensive simulations seem to indicate that ordinary
scaling is recovered asymptotically \cite{Cornell}.

The final case we consider is the system with homogeneous initial
conditions.  The $A$ and $B$ particles are randomly distributed with
equal initial densities $n_0$.  Starting from Eq.~(\ref{eom}) one
can show that the average over initial conditions yields a density
$\overline{\langle a\rangle}\sim t^{-d/4}$ for $d<4$ \cite{LC}.  As
argued by Bramson and Lebowitz, the $A$ and $B$ particles segregate
asymptotically, which may be shown directly from
Eq.\ (\ref{eom}) for $2<d<4$ as follows.  We use the notation
$\langle a\rangle\rightarrow a$, since $R=\Gamma ab$ for $d>2$.

\begin{figure}
\epsfxsize=2.7in
\centerline{\epsfbox{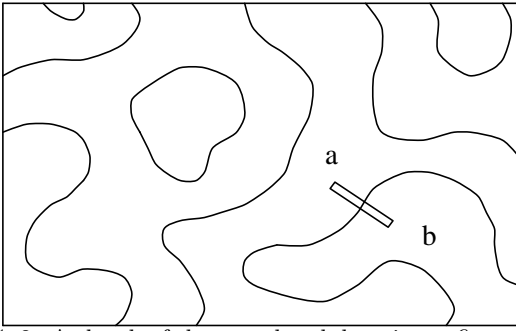}}
\caption{A sketch of the percolated domain configuration in $d=2$.  The
domains have a characteristic length scale which grows in time as
$t^{1/2}$.}\label{f:perc}
\end{figure}

Given that $\overline f\>^2\le\overline{f^2}$, where $f$ is a real
quantity, then
\begin{equation}
\overline{\min(a,b)}\>^2\le{1\over 2}\overline{\bigl(a^2+b^2-(a+b)|a-b|
\bigr).}
\end{equation}
Since $a$ and $b$ correspond to the physical density for a particular initial
condition, they are everywhere non-negative.  Therefore
$a+b\ge|a-b|$ at every point $({\bf x},t)$, which implies that
\begin{equation}
\overline{\min(a,b)}\>^2\le{1\over 2}\overline{\bigl(a^2+b^2-(a-b)^2\bigr)}
=\overline{ab}.
\end{equation}
{}From Eq.\ (\ref{eom}) it follows that
\begin{equation}
\overline{ab}={1\over\Gamma}\partial_t\overline a=O(t^{-1-d/4}),
\end{equation}
so one has the result
\begin{equation}
\overline{\min(a,b)}\le O(t^{-1/2-d/8})\ll t^{-d/4}
\end{equation}
when $d<4$.  Therefore the local minority density goes to zero relative
to that of the majority species.
A sketch of the segregated domains in $d=2$ is shown in Fig.\ \ref{f:perc}.

In this case we may make a similar analogy to the steady-state case, except
now the height at the edge of the depletion region is time-dependent.
The rectangle in Fig.\ \ref{f:perc} gives the cross-section of a reaction
zone, which will have same form as Fig.\ \ref{f:rz}.  The time dependence
of the height is in general
complicated, but we argue that in most regions it will
scale with the bulk density.  The picture one has is that the densities
in the bulk will be fairly uniform since there the particles diffuse
without reacting.  This uniform density must scale as $t^{-d/4}$, so
that averages over the reaction zones of the whole system
will yield  exponents which can be derived simply by taking $a_0,b_0
\rightarrow t^{-d/4}$.  The resulting steady state analog is given
by $J\sim t^{-(d+2)/4}$.  This gives the scaling results Eqs.\ (\ref{scale:a})
and (\ref{scale:R}), with the exponents listed in table \ref{tb} for
the homogeneous case.

The exponents $\alpha$ and $\beta$ are, to our knowledge, new results.
We note that $\alpha\rightarrow 1/2$ from below as $d\rightarrow 4$.  This
is consistent with the view that segregation is breaking down, since the
reaction zones are then on the same scale as the domains.
The exponent $\gamma$ was derived by Leyvraz and Redner \cite{LR,Leyvraz}
by a different method.
They were interested in the quantity $\ell_{AB}$, which is the
characteristic nearest-neighbor distance when the nearest neighbor of a
particle is of the opposite species.  This quantity can be measured
directly in simulations.  Since the particles are segregated, the
contributing particles to the distribution for $\ell_{AB}$ must be within
the reaction zone, and therefore $\ell_{AB}\sim\ell_{\rm rz}\sim t^\gamma$.
Leyvraz and Redner found
from both scaling arguments and simulations values of $\gamma$ which agree
with our predictions for $d=1,2$.  However, they argue for $d>2$ that
$\gamma=1/4$, and show the results of $d=3$ simulations.  Our prediction
for $d=3$ is $\gamma=5/18$, which is very close to $t^{1/4}$.  It appears
from the data of \cite{Leyvraz} that either value of $\gamma$ is an equally
good fit.  A numerical value for a fit to the data is not provided.
The significance of this discrepancy is that the nearest-neighbor distance
in the bulk ${\overline{\langle a\rangle}\>}^{-1/d}\sim t^{1/4}$ for $d<4$.
We claim that $\ell_{\rm rz}$ represents a new length scale for all $d<4$.

A check on the above results is to calculate the characteristic
domain size.  Integrating the equations of motion
(\ref{eom}) over the entire system of size $V$ gives
\begin{equation}
\int d^dx\partial_t\langle a\rangle\sim -\int d^dx\langle ab\rangle
\end{equation}
or equivalently
\begin{equation}
Vt^{-1-d/4}\sim A\int dx_\perp\>R(x_\perp)\sim At^{-\beta+\alpha}
\end{equation}
where $A$ is the interfacial area of the domain boundaries.
This first result follows from the fact that the only significant
contribution to the reaction rate comes from the reaction zones, the
second from Eq.\ (\ref{scale:R}).  This leads to the characteristic
domain size
\begin{equation}
\ell=V/A\sim t^{-\beta+\alpha+1+d/4}\sim t^{1/2}
\end{equation}
for all $d<4$.  This result, while often assumed, follows
for $d<2$ only because $R\neq\langle a\rangle\langle b\rangle$.

We would like to thank Stephen Cornell and Michel Droz for useful
discussions.  This work was supported by a grant from the SERC.

\end{document}